\documentclass[review,a4paper,12pt]{elsarticle}
\usepackage{amssymb}
\usepackage{lineno,hyperref}
\usepackage{epsfig}
\usepackage{epstopdf}
\usepackage{graphicx}

\modulolinenumbers[5]

\journal{Journal of Nucl. Instrum. Meth. B}

\begin{document}

\title{Yield estimation of neutron-rich rare isotopes 
induced by 200 MeV/u $^{132}$Sn beams by using GEANT4}

\author{Jae Won Shin}
\ead{shine8199@skku.edu}
\address{Department of Physics, Sungkyunkwan University,
Suwon 440-746, Korea}
\author{Kyung Joo Min}
\address{Department of Energy Sience, Sungkyunkwan University,
Suwon 440-746, Korea}
\author{Cheolmin Ham}
\address{Department of Energy Sience, Sungkyunkwan University,
Suwon 440-746, Korea}
\author{Tae-Sun Park}
\address{Department of Physics, Sungkyunkwan University,
Suwon 440-746, Korea}
\author{Seung-Woo Hong}
\ead{swhong@skku.ac.kr}
\address{Department of Physics, Sungkyunkwan University,
Suwon 440-746, Korea}

\date{4 March 2015}

\begin{abstract}
A so-called ``two-step reaction scheme", 
in which neutron-rich rare isotopes obtained from 
ISOL are post-accelerated and 
bombarded on a second target, 
is employed to estimate the production yields of 
exotic rare isotopes. 
The production yields of neutron-rich rare isotope fragments 
induced by 200 MeV/u $^{132}$Sn beams 
bombarded on a $^{9}$Be target are estimated
with Monte Carlo code, GEANT4. 
To substantiate the use of GEANT4 for this study, 
benchmark calculations are done for 
80 MeV/u $^{59}$Co, 
95 MeV/u $^{72}$Zn, 
500 MeV/u $^{92}$Mo, 
and 950 MeV/u $^{132}$Sn beams on the $^{9}$Be target. 
It is found that $^{132}$Sn beams 
can produce neutron-rich rare isotopes 
with 45 $\leq$ Z $\leq$ 50 
more effectively than $^{238}$U beams 
at the same energy per nucleon. 
\end{abstract}


\maketitle

\section{Introduction}

The rare isotope beams (RIBs) provide us 
with immense opportunities in wide areas that 
are not covered by stable ion beams, 
which include studies of super-heavy nuclei, 
neutron skins, explosive nucleosynthesis 
and nuclear structure for neutron-rich RIs.
These studies are also closely 
related to questions such as
the origin of elements 
and the fundamental symmetries.
By producing more exotic rare isotopes, 
we may approach new regions of nuclear structure 
and can have a chance to reach 
the limits of nuclear stability 
and previously unexplored state of matter. 
Productions of exotic RIBs are 
being carried out in facilities such as 
ISOLDE at CERN, 
ISAC at TRIUMF, 
RIBF at RIKEN, 
HRIBF at ORNL 
and IGISOL at JYFL, 
and are being planned in 
FRIB at MSU, 
SPIRAL2 at GANIL, 
FAIR at GSI, 
and SPES at INFN. 
The science goals to be explored 
with RIBs are well discussed 
in many reports such as 
U.S. Long Range Plans \cite{plan1}, 
OECD Global Science Forum \cite{plan2}, 
NuPECC Long Range Plan 2010 \cite{plan3} and 
reviews such as Ref. \cite{plan_review}. 

Some of the facilities mentioned above use 
Isotope Separation On Line (ISOL) method \cite{isol1, isol2} 
to produce RIBs, 
while others use 
In-Flight Fragmentation (IFF) method \cite{iff1, iff2}. 
However, as an attempt for producing extremely neutron-rich 
isotopes, a so-called ``two-step reaction scheme" (TSRS) 
has been suggested in Ref. \cite{Sn1}. 
By post-accelerating some of the
fission fragments from ISOL 
and fragmenting them one step further, 
one could produce even more
neutron-rich nuclei covering 
a wider region of elements which would be 
poorly populated by using only either ISOL or IFF method.
It was suggested that 
a long-lived neutron-rich rare isotope such as
$^{132}$Sn produced by 
fission of $^{238}$U induced by proton beams 
could be a good candidate as primary RIBs. 

Recently, TSRS has been experimentally validated 
at the FRagment Separator (FRS) in GSI \cite{Sn2}, 
where they have first 
produced $\sim 10^{3}$ $^{132}$Sn s$^{-1}$ by 
impinging 950 MeV/u $^{238}$U beams on the Pb target 
and bombarded $^{132}$Sn beams on a $^{9}$Be target.
Their results indicate that TSRS 
can be indeed effective in producing 
more exotic medium-mass neutron-rich isotopes.
According to the design summary of 
RAON in Korea \cite{koria1, koria2, koria3}, 
the expected intensity of $^{132}$Sn can be as large as 
about $10^{8}$ particles per second (pps).
It is the unique feature of RAON 
that the isotopes generated by 
ISOL can be post-accelerated 
and injected to the IFF facility 
for producing more exotic RIBs. 

In this work, we have performed 
GEANT4 (GEometry ANd Tracking) \cite{g4n1, g4n2} simulations 
for estimating the production yields of 
exotic rare isotopes by using TSRS.
We have calculated the yields of 
rare isotopes by considering the bombardment of 200 MeV/u $^{132}$Sn 
and $^{238}$U beams on 
the $^{9}$Be target and compared the results.
By comparing the yields from the two different beams, 
we may quantify 
the efficiency of TSRS in producing neutron-rich rare isotopes.
GEANT4 has been used for a number of simulations, 
but has not been used much for nucleus-nucleus (AA) collisions.
For checking the validity of 
the use of GEANT4 for this purpose, 
we have first calculated the isotopic production cross sections for 
AA collisions and compared them with experimental data 
as well as 
with those obtained from another Monte Carlo code, PHITS \cite{phits0}. 
We have also compared the results with 
the predictions from 
other popular empirical models, 
EPAX2 \cite{epax2} and EPAX3 \cite{epax3}. 

The outline of the paper is as follows. 
In Sec. \ref{meth-sec}, 
the simulation methods used in our study are summarized.
In Sec. \ref{result-sec}, 
we first show benchmark calculation results 
and present the production yields of the isotopes 
obtained by $^{132}$Sn and $^{238}$U beams at 200 MeV/u. 
A summary is given in Sec. \ref{sum-sec}.
%

\section{Simulation method
\label{meth-sec}}

Before we use GEANT4 (v10.0) \cite{g4n1, g4n2} for TSRS, 
we have first performed benchmark simulations. 
The production cross sections 
of the isotopes fragmented by AA collisions 
are calculated by using hadronic models such as 
G4BinaryLightIonReaction \cite{BICref1, g4_bicLion} 
and G4QMDReaction \cite{g4qmdRef}. 
We have also performed 
similar simulations with PHITS (v2.52) \cite{phits0} 
by using JAERI Quantum Molecular Dynamics (JQMD)~\cite{niita1} 
for comparison with the results from GEANT4. 

We distinguish our simulation methods by referring to them as 
``G4-BIN", ``G4-QMD" and ``P-JQMD": 
\begin{itemize}
\item{G4-BIN:}
GEANT4 simulation with G4BinaryLightIonReaction model~\cite{BICref1, g4_bicLion}.
\item{G4-QMD:}
GEANT4 simulation with G4QMDReaction model~\cite{g4qmdRef}.
\item{P-JQMD:}
PHITS simulation with JAERI Quantum Molecular Dynamics 
(JQMD)~\cite{niita1}.
\end{itemize}

To see the effectiveness of TSRS, 
the production yields of rare isotopes by bombarding 200 MeV/u $^{132}$Sn 
and $^{238}$U beams on the $^{9}$Be target 
are calculated and compared with each other. 
As our simulation model, 
the G4-BIN is used for the calculations 
because the model gives reliable results 
as will be shown in Sec. \ref{Bench-sec}.
In the production of the isotopes 
through the fragmentation by AA collisions, 
ionization processes cause 
the energy loss of the ions in the target.
The G4ionIonisation is used for simulating such energy losses. 
The above mentioned hadronic and electromagnetic models 
are described on the web \cite{g4_web}, 
and the Physics Reference Manual \cite{g4_physRef} 
is also available.

\section{Results
\label{result-sec}}

\subsection{Benchmark test of simulation methods 
for the production cross sections
\label{Bench-sec}}

AA collisions involve extremely complicated processes,
and it is not yet well established 
how accurately various simulation tools
can describe them.
We thus begin with presenting our benchmarking calculations
for the adopted hadronic models.

Recently, 
several simulations have been performed for AA collisions
with stable beams such as 
$^{12}$C \cite{12C_1, 12C_2, 12C_3}, 
$^{20}$Ne and $^{24}$Mg \cite{24Ne_1}, 
$^{40}$Ar \cite{40Ar_1, 40Ar_56Fe_2}, 
and $^{56}$Fe \cite{40Ar_56Fe_2, 56Fe_1} 
by using Monte Carlo codes 
such as GEANT4, PHITS, FLUKA \cite{fluka1, fluka2} and so on. 
In these studies, 
partial and total fragmentation cross sections have been 
calculated, 
and reasonable agreements 
with the experimental data are obtained.

In this work, we extend such studies by considering
heavier neutron-rich beams including
80 MeV/u $^{59}$Co \cite{bench2}, 
95 MeV/u $^{72}$Zn \cite{bench3}, 
500 MeV/u $^{92}$Mo \cite{bench4}, 
and 950 MeV/u $^{132}$Sn \cite{Sn2} beams. 
Furthermore, in contrast to the above studies 
\cite{12C_1, 12C_2, 12C_3, 24Ne_1, 40Ar_1, 40Ar_56Fe_2, 56Fe_1}, 
in which the production cross sections 
are calculated for each element with a sum over isotopes, 
we analyze the `isotopic' production cross sections of 
AA collisions for detailed comparisons. 

\begin{figure}[tbp]
\begin{center}
\epsfig{file=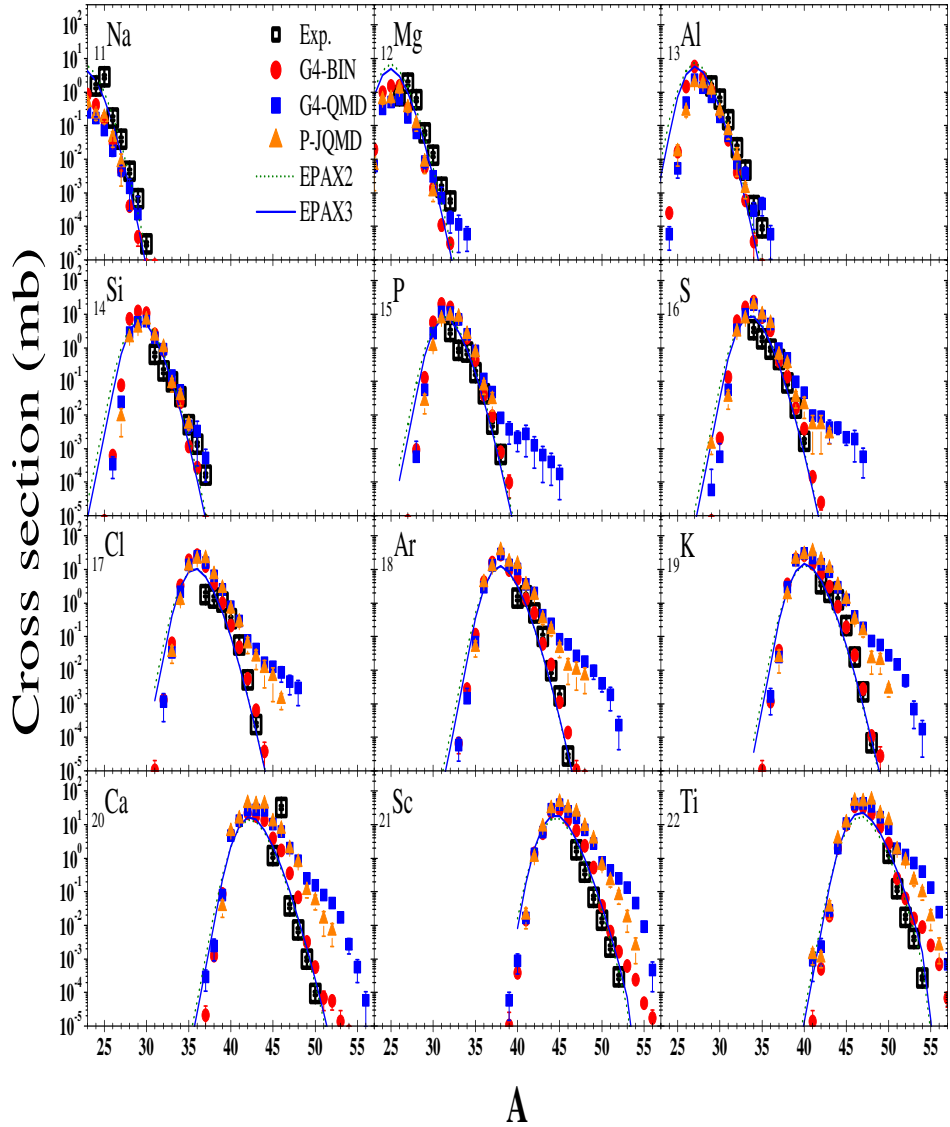, height=7in, width=5.5in}
\end{center}
\caption{(Color online) 
Production cross sections of 
various isotopes due to the collision 
$^{59}$Co + $^{9}$Be at 80 MeV/u.
The open squares in black denote the experimental data \cite{bench2},
and the dashed (solid) lines represent EPAX2 (EPAX3) results.
The simulation results are denoted by the filled symbols; 
the red circles for G4-BIN, 
the blue squares for G4-QMD
and the orange triangles for P-JQMD.
}
\label{Fig1}
\end{figure}
\begin{figure}[tbp]
\begin{center}
\epsfig{file=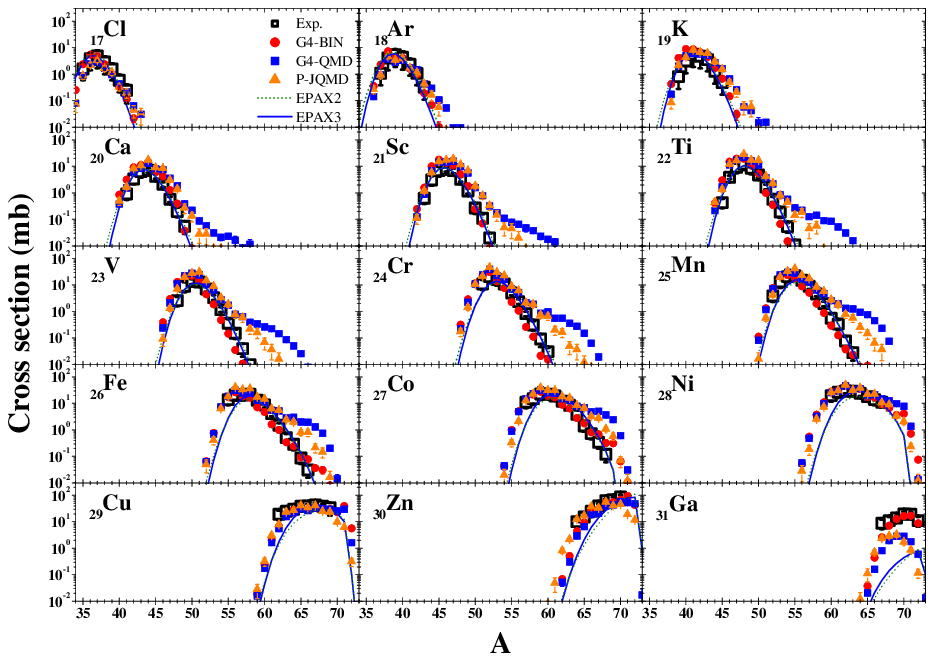, height=7in, width=5.5in}
\end{center}
\caption{(Color online) Production cross sections of 
various isotopes due to the collision 
$^{72}$Zn + $^{9}$Be at 95 MeV/u. 
The open squares in black denote the experimental data \cite{bench3}.
See the caption of Fig. \ref{Fig1} for the meaning of other symbols.}
\label{Fig2}
\end{figure}
\begin{figure}[tbp]
\begin{center}
\epsfig{file=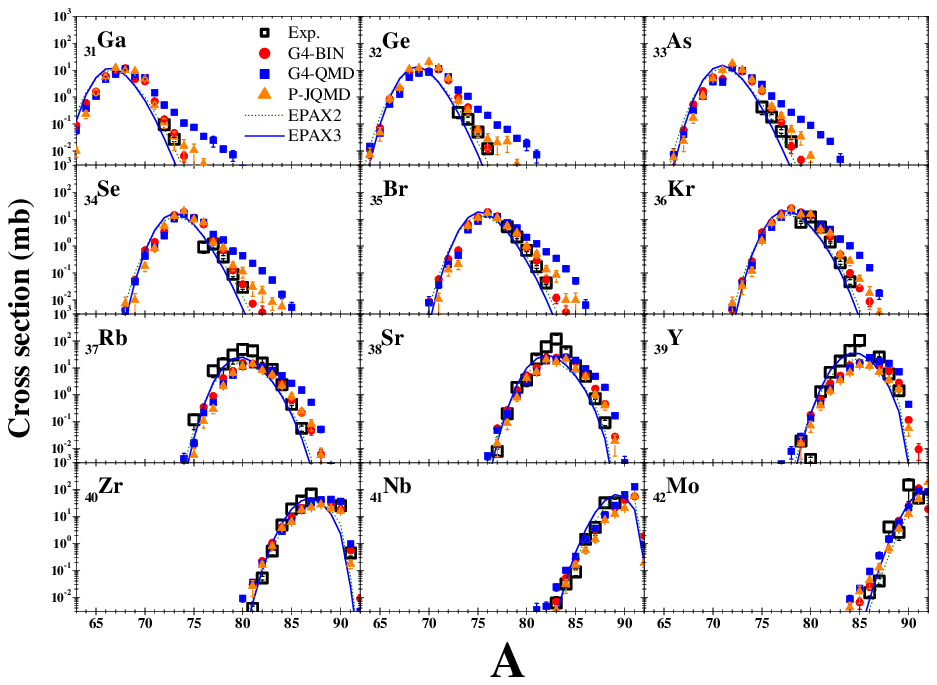, height=7in, width=5.5in}
\end{center}
\caption{(Color online) Production cross sections of 
various isotopes due to the collision 
$^{92}$Mo + $^{9}$Be at 500 MeV/u.
The open squares in black denote the experimental data \cite{bench4}.
See the caption of Fig. \ref{Fig1} for the meaning of other symbols.}
\label{Fig3}
\end{figure}
\begin{figure}[tbp]
\begin{center}
\epsfig{file=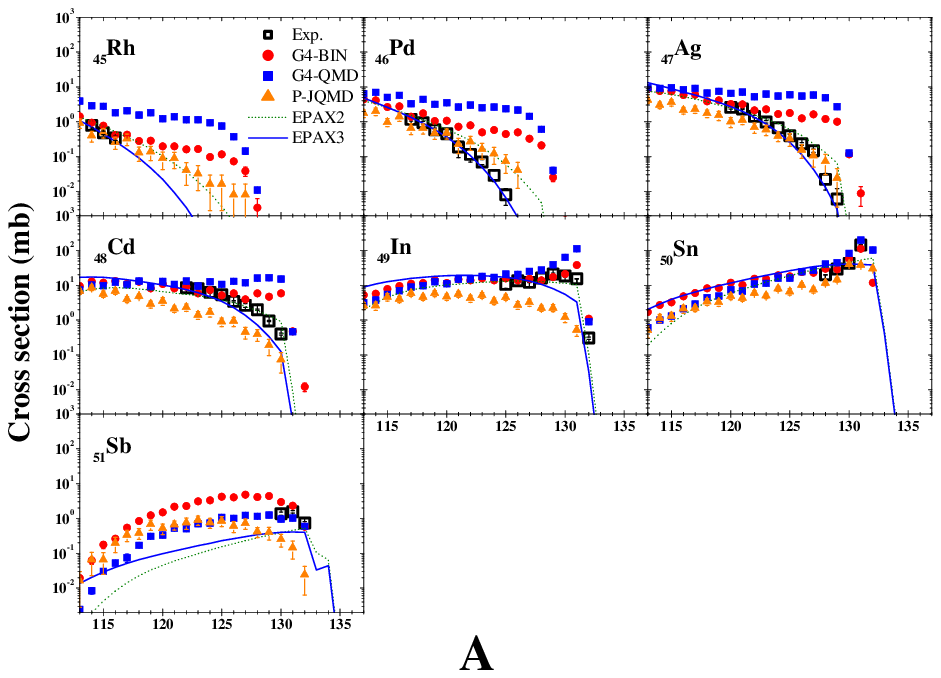, height=7in, width=5.5in}
\end{center}
\caption{(Color online) Production cross sections of 
various isotopes due to the collision 
$^{132}$Sn + $^{9}$Be at 950 MeV/u.
The open squares in black denote the experimental data \cite{Sn2}.
See the caption of Fig. \ref{Fig1} for the meaning of other symbols.}
\label{Fig4}
\end{figure}

Our calculated production cross sections 
for the above reactions
are shown in Figs. \ref{Fig1} $\sim$ \ref{Fig4} 
by the filled symbols;
the red circles for G4-BIN, 
the blue squares for G4-QMD 
and the orange triangles for P-JQMD.
For comparison,
we have also plotted the predictions from 
EPAX2 \cite{epax2} (the dashed lines) and 
EPAX3 \cite{epax3} (the solid lines) 
together with 
the experimental data (the open squares in black) 
taken from the EXFOR database \cite{exfor}. 

Figures \ref{Fig1}$\sim$\ref{Fig3} 
show the results for 
80 MeV/u $^{59}$Co, 
95 MeV/u $^{72}$Zn, 
and
500 MeV/u $^{92}$Mo beams, respectively. 
The G4-BIN (red circles)
is found to be superior to other models
in reproducing the experimental data.
EPAX2 and EPAX3 
also reproduce well the experimental data for most cases,
but underestimate in some cases,
in particular, for the production of the ${}_{31}$Ga isotopes 
for ${}^{72}$Zn beams.
P-JQMD (orange triangles)
and G4-QMD (blue squares)
overestimate the production of neutron-rich isotopes 
of heavier elements by a factor of $\lesssim 10^4$, 
while their results for light elements are 
as good as those from G4-BIN. 

\begin{figure}[tbp]
\begin{center}
\epsfig{file=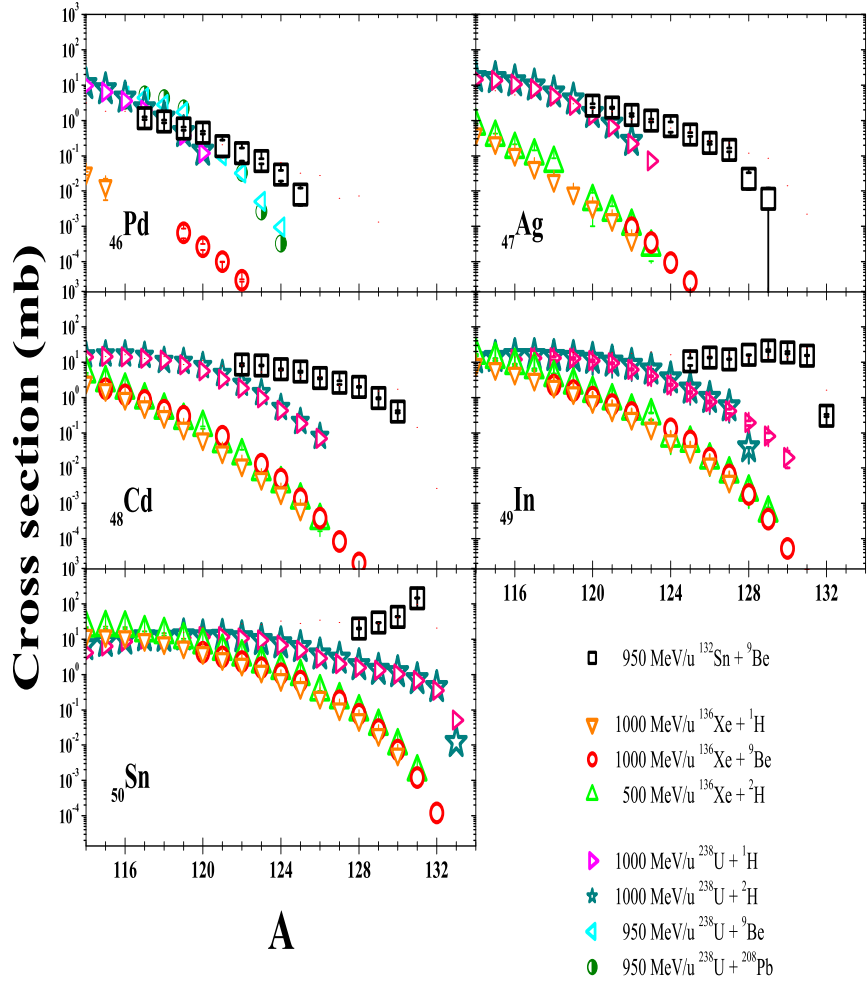, height=6in, width=5in}
\end{center}
\caption{(Color online) 
The experimental production cross sections of 
$_{46}$Pd $\sim$ $_{50}$Sn isotopes.
The black boxes represent 
the production cross sections from 
950 MeV/u $^{132}$Sn + $^{9}$Be \cite{Sn2}.
The orange inverted triangles are those from 
1000 MeV/u $^{136}$Xe + $^{1}$H \cite{136Xe_1}, 
the red circles from 1000 MeV/u $^{136}$Xe + $^{9}$Be \cite{136Xe_2}, and 
the green triangles from 500 MeV/u $^{136}$Xe + $^{2}$H \cite{136Xe_3}.
The pink triangles are from 
1000 MeV/u $^{238}$U + $^{1}$H \cite{238U_1}, 
the blue stars from 1000 MeV/u $^{238}$U + $^{2}$H \cite{238U_2}, 
the blue triangles from 950 MeV/u $^{238}$U + $^{9}$Be \cite{238U_3}, and 
the green circles from 950 MeV/u $^{238}$U + $^{208}$Pb \cite{238U_3}.
} 
\label{Fig5}
\end{figure}

In Fig. \ref{Fig4}, 
the results for $^{132}$Sn + $^{9}$Be reactions 
at 950 MeV/u are shown. 
None of the models can reproduce all the experimental data satisfactorily.
P-JQMD underestimates the production of 
$_{48}$Cd $\sim$ $_{51}$Sb isotopes, 
while G4-QMD overestimates the production of 
$_{45}$Rh $\sim$ $_{49}$In isotopes
by a factor of $\lesssim$ 10$^{3}$. 
On the other hand, EPAX2 and EPAX3 
show rather good overall agreements, 
though EPAX2 overestimates the production of $_{46}$Pd isotopes,
EPAX3 underestimates neutron-rich isotopes of $_{49}$In,
and both underestimate $_{51}$Sb isotopes.
Finally, 
G4-BIN, which shows 
the best performance for other cases, 
is also found to overestimate 
the productions of $_{46}$Pd, $_{47}$Ag, and $_{48}$Cd isotopes
by $\lesssim 10^2$.
Such overestimation of the production cross sections 
needs to be ultimately 
resolved by improving the hadronic models, 
but it is beyond the scope of this work.

These benchmark calculations show that G4-BIN 
gives the best results among the three models.  
Although the benchmark calculations are done for 950 MeV/u $^{132}$Sn beams,
it is expected that G4-BIN would work better than other models
for 200 MeV/u $^{132}$Sn beams as well for the following reason.  
In Fig.~\ref{Fig5} we have plotted 
experimental production cross sections
of $_{46}$Pd $\sim$ $_{50}$Sn isotopes for 
$^{132}$Sn \cite{Sn2}, 
$^{136}$Xe \cite{136Xe_3, 136Xe_1, 136Xe_2} and 
$^{238}$U \cite{238U_1, 238U_2, 238U_3} beams. 
The experimental cross sections are
nearly independent of the incident beam energies 
\cite{epax3, 12C_2, bench3, epax1, Edep_1, Edep_2, Edep_3, 136Xe_3}.
It is clearly seen that 
the cross sections are mainly characterized by 
the species of the incident beams, 
and the energy-dependence is rather weak.

\begin{figure}[tbp]
\begin{center}
\epsfig{file=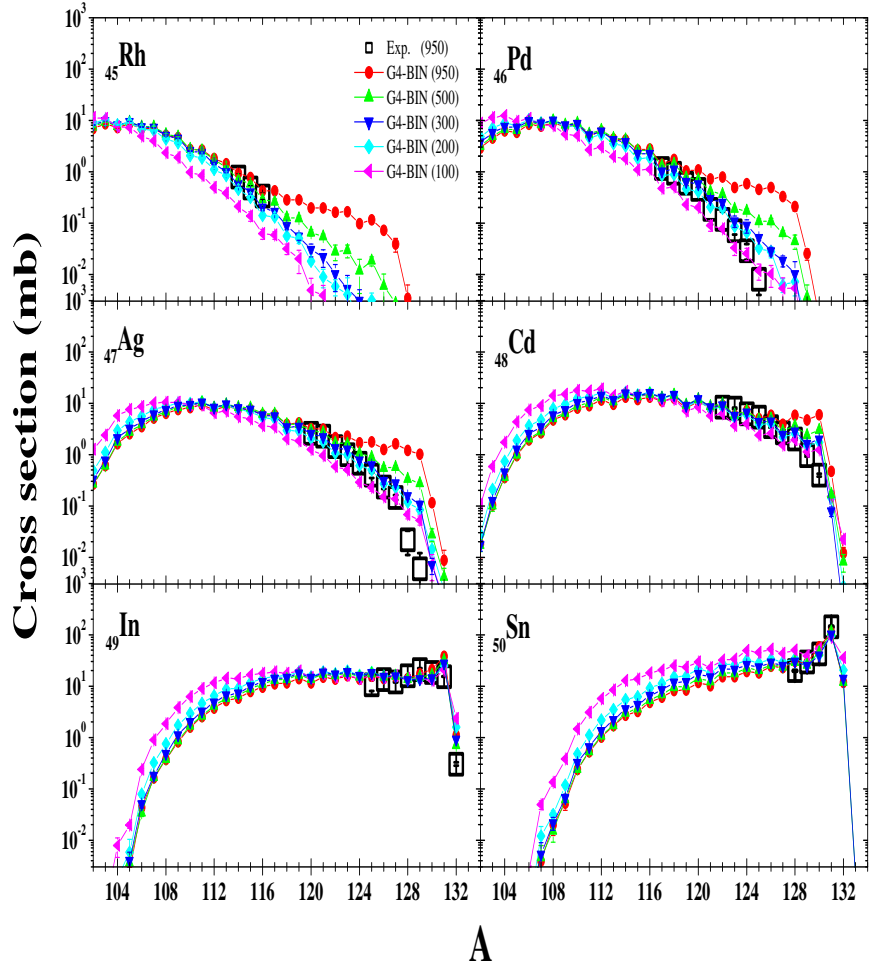, height=6.in, width=5.in}
\end{center}
\caption{(Color online) Production cross sections of
$^{132}$Sn + $^{9}$Be.
The open squares in black denote the experimental data 
at 950 MeV/u \cite{Sn2},
and the filled symbols with lines represent 
the G4-BIN results
for a few selected incident energies.
} 
\label{Fig6}
\end{figure}

We plot in Fig. \ref{Fig6} G4-BIN predictions for
the production cross sections of 
$_{45}$Rh $\sim$ $_{50}$Sn isotopes 
due to the $^{132}$Sn beams 
at 100, 200, 300, 500 and 950 MeV/u.
The G4-BIN predictions at the energy of $\sim$ 200 MeV/u 
agree well with the 950 MeV/u data 
regardless of the atomic number Z.
If the aforementioned experimentally 
observed energy-independence 
of the cross section still holds down to 200 MeV/u, 
we expect a good agreement will be obtained 
between the data and the G4-BIN predictions at 200 MeV/u, 
for which the beam energy 
is being planned for RAON \cite{koria1, koria2, koria3}.
In fact, G4-BIN shows a small energy-dependence.
While the calculated yields induced by 200 MeV/u $^{132}$Sn beams
agree with the experimental data obtained at 950 MeV/u, 
G4-BIN prediction made at 950 MeV/u overestimates the data 
obtained at 950 MeV/u for $_{46}$Pd, $_{47}$Ag, and $_{48}$Cd isotopes
with larger A. 
The energy-dependence of the G4-BIN is weak for 
Z$=$ 49 and 50, but becomes noticeable as Z decreases below 49.

\begin{figure}[tbp]
\begin{center}
\epsfig{file=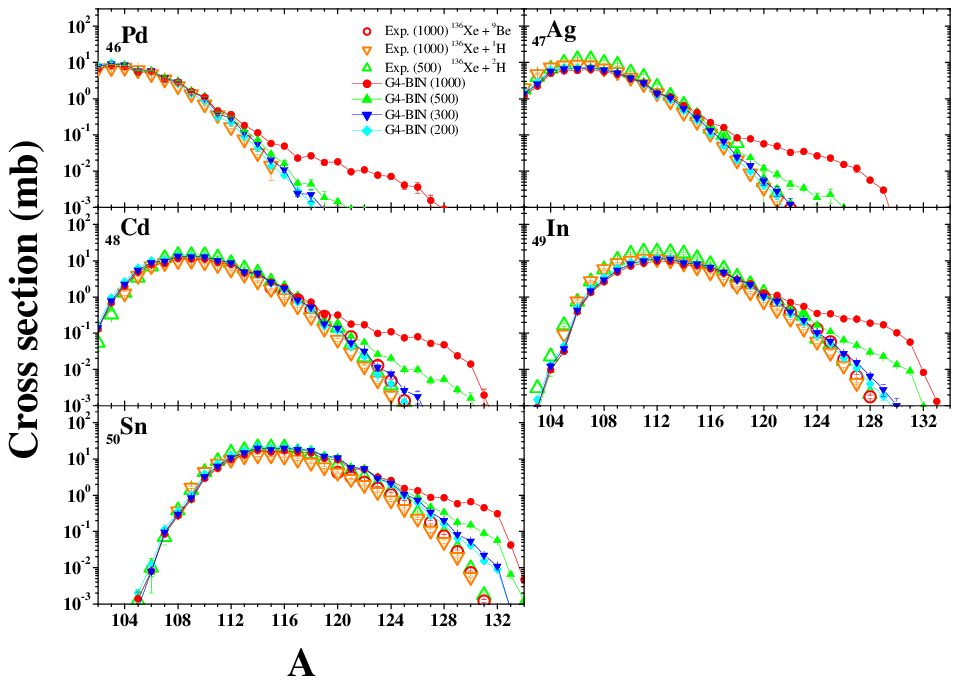, height=6.in, width=6.in}
\end{center}
\caption{(Color online) 
Production cross sections of $^{136}$Xe beams.
The red circles are for 1000 MeV/u $^{136}$Xe + $^{9}$Be \cite{136Xe_2}, 
the inverted orange triangles are for 
1000 MeV/u $^{136}$Xe + $^{1}$H \cite{136Xe_1}, and 
the green triangles for 500 MeV/u $^{136}$Xe + $^{2}$H \cite{136Xe_3}.
The filled symbols with lines denote 
the G4-BIN results for a few selected values of incident energy.
}
\label{Fig7}
\end{figure}

Similar features can be seen for $^{136}$Xe beams,
whose experimental data are available at 500 and 1000 MeV/u.
As shown in Fig. \ref{Fig7}, 
the experimental data are rather independent of energy.
The predicted cross sections at 200 MeV/u
are close to the data obtained at 500 and 1000 MeV/u,
and the calculated cross sections show some energy dependence 
for isotopes with larger A, which is a feature observed 
for $^{132}$Sn beams as well.

From the above comparisons of 
the results by using different models, 
it is found that the G4-BIN gives most reasonable results 
for the isotopic productions among the three models. 
The G4-QMD does not reproduce well the experimental data 
especially for the production of neutron-rich isotopes. 
The P-JQMD either overestimates or 
underestimates the experimental data.
Thus, 
we have chosen the G4-BIN as our simulation model,
and the results presented in the next subsection 
are calculated by the G4-BIN.

\subsection{Production yields of neutron rich isotopes by using TSRS}

\begin{figure}[tbp]
\begin{center}
\epsfig{file=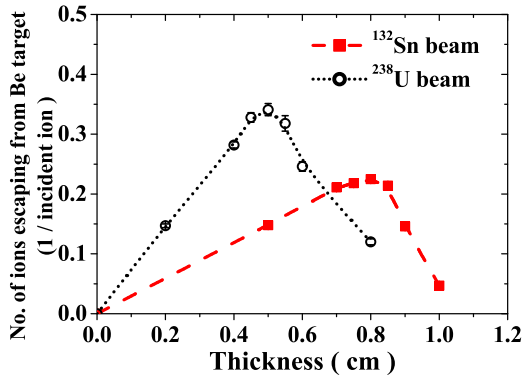, width=3in}
\end{center}
\caption{(Color online) 
The total number of ions per incident ion escaping from the Be target.
The red squares and the open black circles denote the results from 
$^{132}$Sn and $^{238}$U beams of 200 MeV/u, respectively. 
The dotted and the dashed lines are to guide the eyes.}
\label{Fig8}
\end{figure}

To see the effectiveness of TSRS, 
one needs to estimate the production yields from TSRS. 
Production yields are sensitive to the target thickness.
As the target thickness increases, 
the number of the collisions 
between the incident ions and 
the target nuclei also increases.
But if the target is too thick, 
it becomes hard for the generated nuclides 
to escape from the target.
Figure \ref{Fig8} shows 
the dependence of the yields on the 
target thicknesses. 
It can be seen that at the energy of 200 MeV/u the optimal thickness
is 0.8 cm and 0.5 cm for $^{132}$Sn and $^{238}$U beams,
respectively,
and thus these thicknesses are chosen
for our simulations 
of the production yields.

\begin{figure}[tbp]
\begin{center}
\epsfig{file=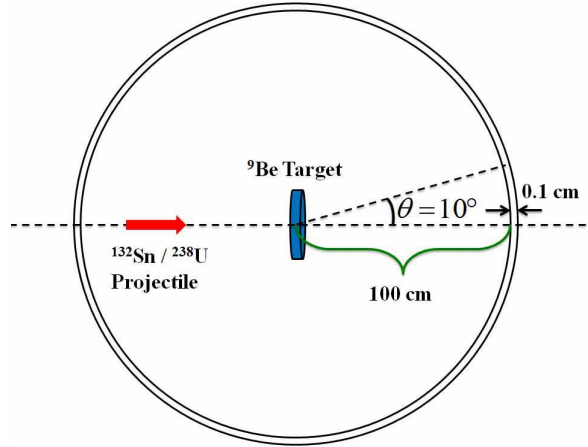, width=3in}
\end{center}
\caption{Schematic diagram of the simulation setup.}
\label{Fig9}
\end{figure}

A schematic geometrical setup of our simulation 
is depicted in Fig. \ref{Fig9}.
Incident ions are treated as a pencil beam.
The $^{9}$Be target is modeled 
as a cylinder of radius 1 cm, 
whose thickness is chosen as 0.8 cm and 0.5 cm 
for $^{132}$Sn and $^{238}$U beams, respectively.
The scoring region of a spherical shell shape 
is placed surrounding the target.
The inner and the outer radii of the scoring 
region are chosen to be 100 cm and 100.1 cm, respectively.
The target area is in vacuum, and 
the thickness of the scoring region 
is chosen as 0.1 cm arbitrarily for convenience 
and does not affect the results.
We score only those isotopes that reach the scoring region. 
The scoring shell consists of 18 angular bins 
with an interval of 10 degrees.
We find that about $\sim$98\% of the yields 
are concentrated in the 
forward direction with ${\theta} <$ 10$^{\circ}$ 
and thus we show only the 
scoring results in the bin with ${\theta} <$ 10$^{\circ}$ 
for convenience. 

\begin{figure}[tbp]
\begin{center}
\epsfig{file=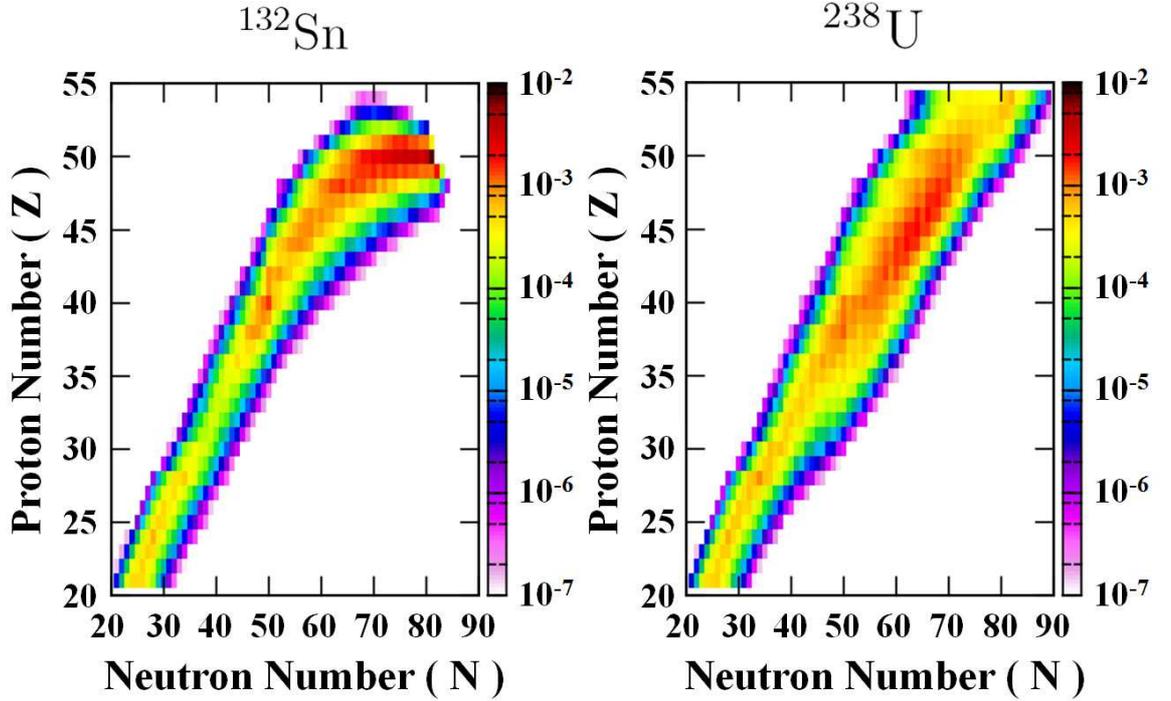, width=6in}
\end{center}
\caption{(Color online) The isotopic production yields
plotted in the N-Z plane,
for 200 MeV/u $^{132}$Sn (left) 
and $^{238}$U (right) beams on the $^{9}$Be target.
}
\label{Fig10}
\end{figure}

\begin{figure}[tbp]
\begin{center}
\epsfig{file=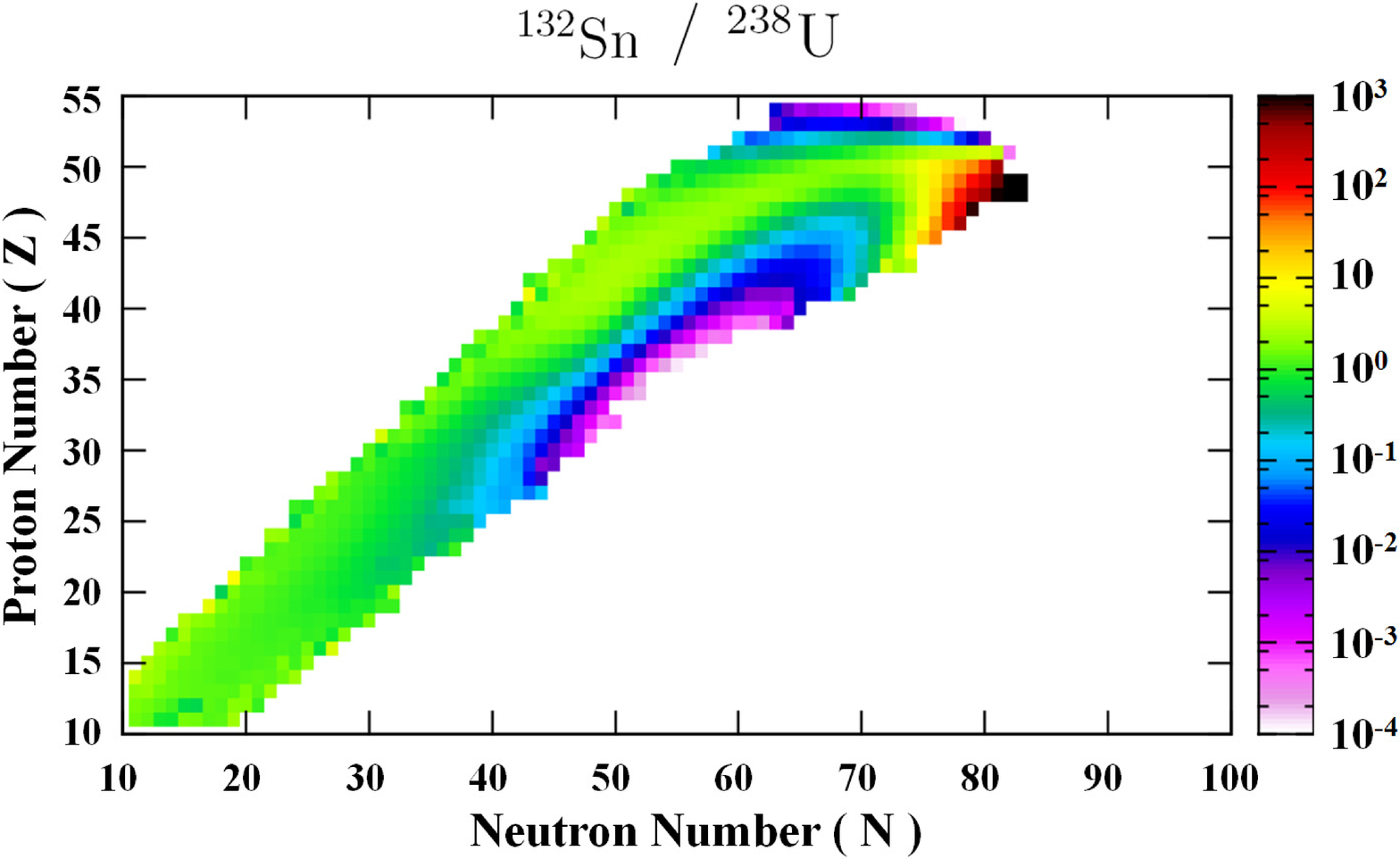, width=6in}
\epsfig{file=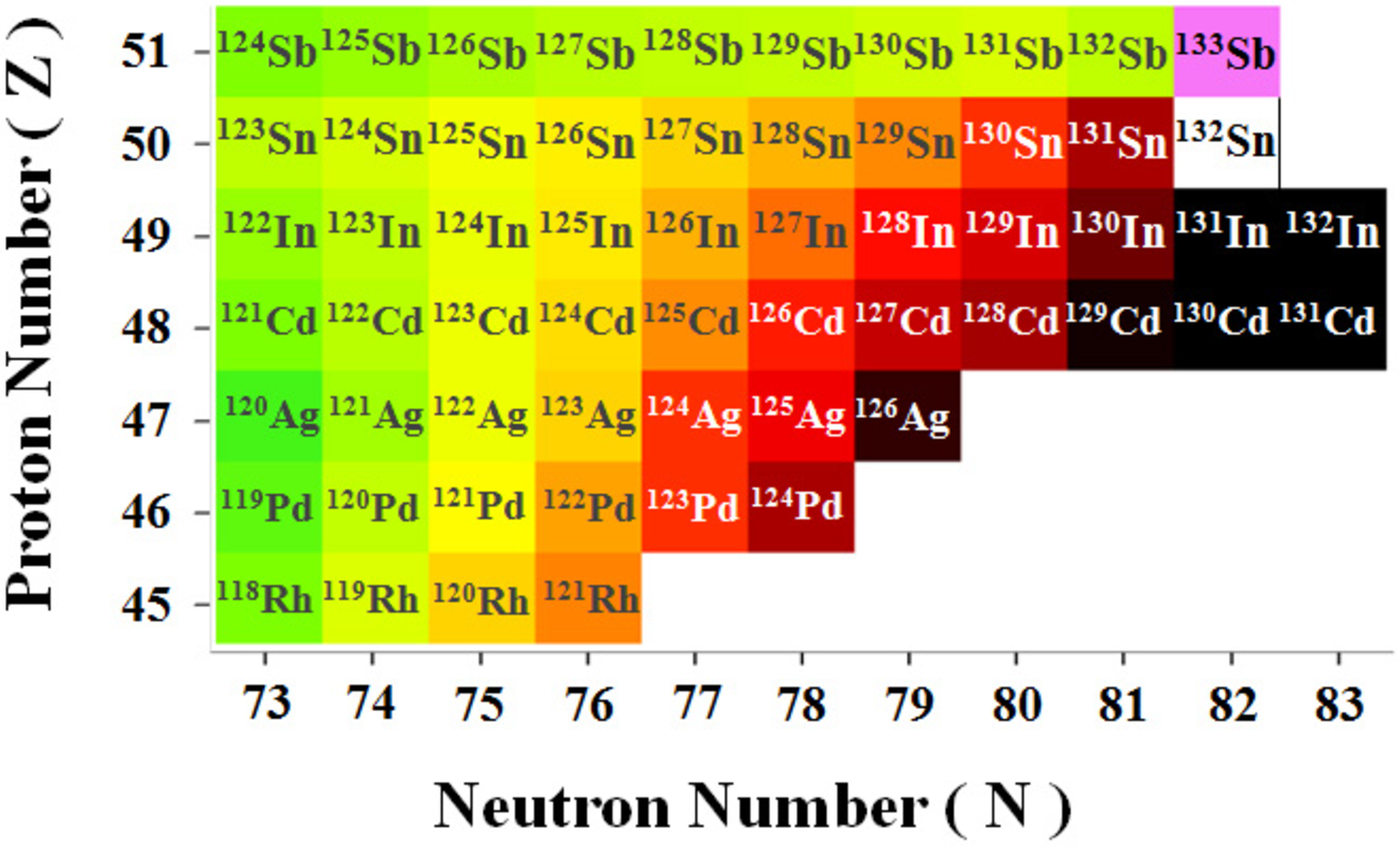, width=4in}
\end{center}
\caption{(Color online) 
The ratios of the yields due to $^{132}$Sn beams to 
the yields due to $^{238}$U beams. 
The region where the ratio is greater than 10$^{3}$ is denoted by black.}
\label{Fig11}
\end{figure}

Figure \ref{Fig10} shows the yields of the nuclides 
produced by 200 MeV/u $^{132}$Sn and $^{238}$U beams.
$^{132}$Sn beams produce isotopes mainly with 
45 $\leq$ Z $\leq$ 50,
while the products induced by $^{238}$U beams
are concentrated in the diagonal region of the N-Z plane
that mainly consists of stable nuclides.
We plot the ratios of the yields of the nuclides 
produced by $^{132}$Sn to those produced by $^{238}$U 
in Fig. \ref{Fig11},
which clearly shows that $^{132}$Sn 
can produce more neutron-rich isotopes 
of 45 $\leq$ Z $\leq$ 50 by nearly 
two or three orders of magnitudes 
than $^{238}$U beams.


\begin{figure}[tbp]
\begin{center}
\epsfig{file=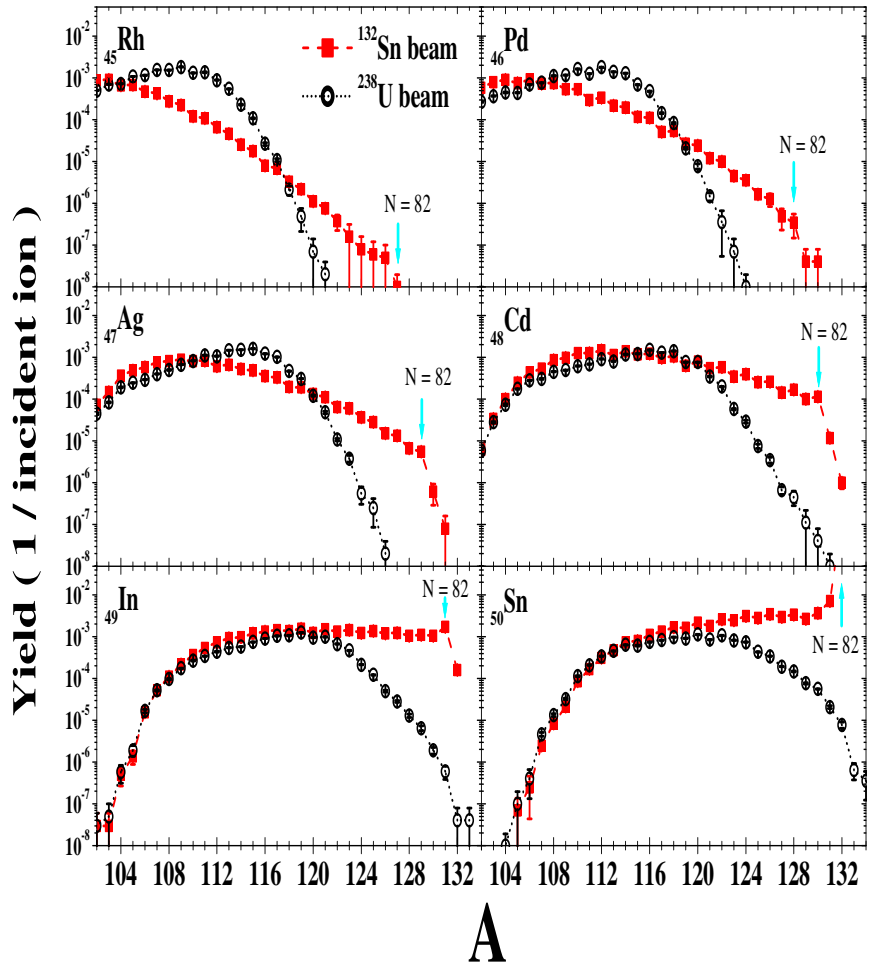, height=6.in, width=5.in}
\end{center}
\caption{(Color online) 
Yields of ${}_{45}\mbox{Rh}\sim {}_{50}\mbox{Sn}$ isotopes 
with respect to mass number A.
The red squares and the open black circles denote the yields induced by
$^{132}$Sn and $^{238}$U beams, respectively.}
\label{Fig12}
\end{figure}

Detailed production yields of nuclides 
are shown in Fig. \ref{Fig12},
where
the yields
of ${}_{45}\mbox{Rh}\sim {}_{50}\mbox{Sn}$ isotopes
due to $^{132}$Sn and $^{238}$U beams 
are plotted with respect to the mass number A.
The yield curves due to $^{238}$U beams 
are more or less parabolic, 
having the maximum yield value ($\sim 10^{-3}$ per incident ion)
at around A $\simeq$ 2$\times$Z + 20 
for all the Z values considered here. 
As Z increases,
the curves shift to the region of higher A, 
but their shapes are almost 
unchanged for different elements. 
This behavior is in accordance with
the right panel of Fig. \ref{Fig10},
which shows that
the population of the fission fragments of $^{238}$U
is concentrated in the diagonal region in the N-Z plane.

The yield curves due to $^{132}$Sn beams also show the tendency 
to shift to higher A values as Z increases, 
and are similar to those of $^{238}$U in the light mass region
for most elements. 
But the curves deviate from those of $^{238}$U
in the heavier mass region. 
The tails of the yield curves for $^{132}$Sn beams in the larger A region  
do not drop as rapidly as those for $^{238}$U.
This shows that $^{132}$Sn can produce 
more effectively neutron-rich isotopes than $^{238}$U beams.
This tendency becomes stronger as Z increases,
and the tail for $_{49}$In becomes almost flat up to N=82.
The mass region where 
the yields due to $^{132}$Sn exceed those of $^{238}$U 
keeps increasing in general with higher Z: 
A$\ge$118 for $_{45}$Rh, 
A$\ge$119 for $_{46}$Pd, 
A$\ge$120 for $_{47}$Ag, 
A$\ge$122 for $_{48}$Cd, 
A$\ge$123 for $_{49}$In,
and
A$\ge$114 for $_{50}$Sn.

Note that the yield curves due to 
$^{132}$Sn beams show a shell structure 
with the cusps developed 
at the magic number N=82.
The yield for N=82 isotope is about
$10^{-8}$ for $_{45}^{127}$Rh, 
$10^{-6}$ for $_{46}^{128}$Pd, 
$10^{-5}$ for $_{47}^{129}$Ag, 
$10^{-4}$ for $_{48}^{130}$Cd
and $2\times 10^{-3}$ for $_{49}^{131}$In 
in the unit of 1$/$(incident ion).
For all the cases considered,
the yields for the isotopes with N$>$82, 
which means more neutrons than the incident $^{132}$Sn beams, 
are found to be suppressed by a factor of $\simeq 10$ 
for each additional neutron. 

Our results show that 
at the energy of 200 MeV/u 
$^{132}$Sn beams is more effective 
than $^{238}$U beams
in producing neutron-rich isotopes. 
In particular, the yields are enhanced by factors of $10^3$
for $^{131,132}$In, $^{129,130,131}$Cd and $^{127}$Ag.

\section{Summary
\label{sum-sec}}

We have conducted a simulation study 
for the isotope production yields
induced by 200 MeV/u $^{132}$Sn and 
$^{238}$U beams on the $^{9}$Be target
by using G4BinaryLightIonReaction (G4-BIN) hadronic model
provided by GEANT4 as our main engine 
after considering three simulation methods 
such as G4-BIN, G4-QMD and P-JQMD. 
For the production of neutron-rich isotopes 
with 45 $\leq$ Z $\leq$ 50, 
$^{132}$Sn is found to be 
much more efficient than $^{238}$U beams, 
supporting the usefulness of the TSRS mechanism.

In our benchmark calculations for the production cross sections 
to check the accuracy of hadronic models,
however, we have observed that
the G4-BIN has overestimated 
the production of neutron-rich isotopes 
for $_{46}$Pd, $_{47}$Ag, and $_{48}$Cd isotopes
with 950 MeV/u $^{132}$Sn beams. 
As discussed in Sec. \ref{Bench-sec},
the experimental production cross sections 
of $_{46}$Pd $\sim$ $_{50}$Sn isotopes induced by
$^{132}$Sn \cite{Sn2}, 
$^{136}$Xe \cite{136Xe_3, 136Xe_1, 136Xe_2} and 
$^{238}$U \cite{238U_1, 238U_2, 238U_3} beams 
are quite energy-independent, while the calculated
production cross sections have some energy dependence.
Thus, our simulation results with 200 MeV/u beams
is expected to have some uncertainty, though it seems small 
and remains to be resolved. 

\section*{Acknowledgement}
This work was supported in part by the Basic
Science Research Program through the Korea Research
Foundation (NRF-2013R1A1A2063824, NRF-2012R1A1A2007826, NRF-2012M2B2A4030183). 

\bibliography{mybibfile}

%
\end{document}